\documentclass[10pt,letterpaper]{article}
\usepackage{opex3}
\usepackage[]{graphicx}
\usepackage{dcolumn}
\usepackage{bm,multirow}

\begin{document}

\title{Second-order nonlinear mixing in planar photonic crystal microcavities }

\author{Murray W. McCutcheon, Georg W. Rieger, and Jeff F. Young}
\address{Department of Physics and Astronomy, University of British Columbia,
Vancouver, Canada, V6T 1Z1}
\author{Dan Dalacu, Simon Fr\'{e}d\'{e}rick$^*$, Philip J. Poole, and Robin~L.~Williams$^*$}
\address{Institute for Microstructural Sciences, National Research Council, Ottawa, Canada, K1A OR6\\ $^*$\emph{Also at: }Physics Department, University of Ottawa, Ottawa, Canada, K1N 6N5}
\email{murray@phas.ubc.ca}

\begin{abstract}  Second-harmonic and sum-frequency mixing phenomena associated with 3D-localized 
photonic modes are studied in InP-based planar photonic crystal microcavities excited by 
short-pulse radiation near 1550 nm.  Three-missing-hole microcavities that support two 
closely-spaced modes exhibit rich second-order scattering spectra that reflect intra- 
and inter-mode mixing via the bulk InP $\chi^{(2)}$ during ring-down after excitation by the 
broadband, resonant pulse.  Simultaneous excitation with a non-resonant source results 
in tunable second-order radiation from the microcavity. 
\end{abstract}



\section{Introduction}

Microfabricated structures in semiconductor thin films offer the opportunity to tightly 
confine light in nonlinear, transparent media~\cite{Sakoda96,Soljacic, Cowan05, Barclay05}. 
Various micro-disk and micro-toroid structures have been fabricated in semiconductor
membranes to act as ultrasmall optical cavities that support discrete microcavity modes with 
very high quality (Q) factors in excess of $10^8$~\cite{Armani}.  These exhibit optical 
bistability and other nonlinear responses at exceedingly low optical powers when coupled 
efficiently to single mode waveguides~\cite{Kippenberg04, Johnson}.  Engineered defect states 
within planar photonic crystals (PPCs) offer an alternative to micro-disk/toroid-based 
cavities: currently they have not been produced with as high Q values, but their mode 
volumes ($V_m$) are smaller, so that the ratio of their Q values to their mode 
volumes are comparable, or even larger.  $Q/V$ or $Q/\sqrt{V_m}$ are figures of merit 
for cavity-based quantum electrodynamic (cavity QED) phenomena, which provides yet 
another motivation for developing ultrasmall nonlinear structures~\cite{Vahala, Kimble}. 
Enhanced second-harmonic generation (SHG) has been observed in 1D photonic crystal 
microcavities with dielectric~\cite{Trull} and mesoporous silicon
Bragg mirrors~\cite{Dolgova}.  Previous work studying SHG in PPC slabs
showed that when one or both of the fundamental and second-harmonic beams are 
mode-matched to leaky modes of the structure, there is a significant resonant 
enhancement of the
radiated second harmonic~\cite{Cowan02,Mondia}.  This effect was shown to be
due to the local field enhancements associated with the incoming and outgoing resonances.  

In this letter, we report on the second-order response of the 3D-localized states of a 
PPC microcavity in a sub-wavelength thick semiconductor slab that is locally 
excited by a diffraction-limited beam {\em incident perpendicular to the slab}.  
The mode energies, and hence the second-order radiation energies, can be controlled by 
tailoring the photonic crystal defect lattice that defines the microcavity~\cite{PainterPRB}.

\section{Experiment}
\begin{figure}[h]
\centering
\includegraphics[width=10cm]{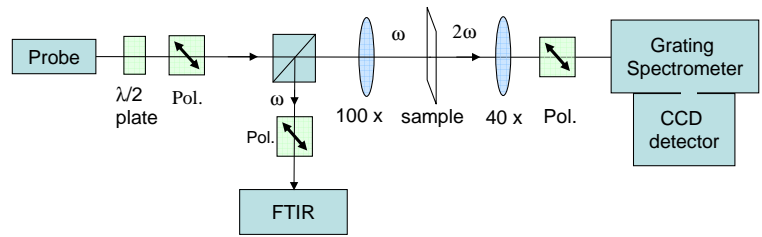}
\caption{\label{fig:setup}Schematic of the experimental set-up.  The linear spectra
are obtained from the reflected resonantly scattered radiation, and detected in
the cross-polarization using a Fourier transform infrared (FTIR) spectrometer.  
The second-order radiation is detected (simultaneously) 
in a transmission geometry.}
\end{figure}

The slab which hosts the PPC microcavity is a 230nm thick [001]-oriented InP 
free-standing membrane mounted on a glass substrate.  It is excited by a train of pulses
from a 80 MHz optical parametric oscillator (OPO) (Spectra Physics) pumped
at 810 nm and focussed through a 100x microscope objective lens~\cite{McCutcheon_APL}.  
As shown in the optical set-up in Figure~\ref{fig:setup},
the resonantly scattered light is collected in reflection, and detected in the 
cross-polarization with respect to the incident beam
using a Bomem Fourier transform infrared spectrometer.  The second-order radiation 
is collected in transmission using a 40$\times$ (NA=0.65) 
microscope objective, and detected using a grating spectrometer and a liquid-nitrogen 
cooled CCD detector.

\section{Results}

\begin{figure}[h]
\centering
\includegraphics[width=10cm]{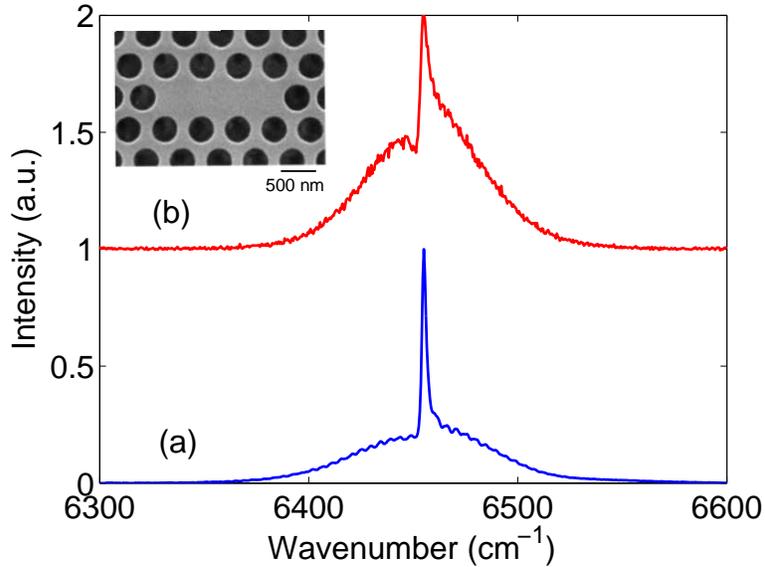}
\caption{\label{fig:highQ}(a) Linear and (b) second-order (plotted at half the energy)
spectra from the high Q mode of an InP 3-h microcavity.  An SEM image of the microcavity
is shown in the inset.}
\end{figure}

The linear resonant scattering spectrum from a three-missing-hole (3h) cavity which 
supports a single mode is shown in Figure~\ref{fig:highQ}(a).  The non-resonant 
background, which has the shape of the excitation spectrum, is radiated by polarization 
driven by the OPO source while it directly interacts with the thin slab.  The resonant 
feature is due to lingering polarization induced by the electric field scattered into 
the resonant mode of the cavity as it ``rings down''~\cite{McCutcheon_APL}.   
When the spectrometer is set to twice the 
frequency, the resulting SHG spectrum (b) closely mimics the linear spectrum.  The 
broad peak is also observed when the untextured InP slab is irradiated by the same beam.  
This {\em non-resonant} second-order scattering corresponds to the second-order 
interaction between the laser pulse and the InP slab.  The sharp feature at exactly 
twice the mode frequency corresponds to the second-order polarization induced by the 
mode, as it rings down, via the $\chi^{(2)}$ of the InP slab.

\begin{figure}[h]
\centering
\includegraphics[width=10cm]{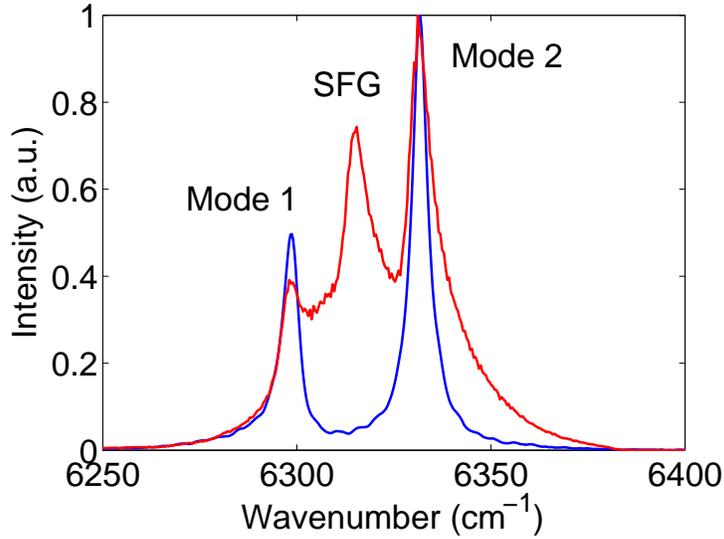}
\caption{\label{fig:2mode}Linear (blue) and nonlinear (red) resonant scattering from a
two-mode 3-h microcavity.  The nonlinear spectrum is plotted at half the energy,
and shows an additional feature due to sum-frequency generation (SFG) 
of the modes.}  
\end{figure}
When a microcavity supports two modes, the second-order spectra are more complex.
Spectra from a 3h-cavity with two closely-spaced modes are shown in 
Figure~\ref{fig:2mode}.  The linear scattering spectrum (blue) shows two resonant features at
6300 cm$^{-1}$ and 6330 cm$^{-1}$, superimposed on a non-resonant background from the 
scattered laser spectrum used to excite the sample.  In the second-order spectrum (red), which
is plotted at half the energy, the lowest and highest energy peaks are at exactly 
twice the frequencies of the microcavity modes evident in the linear spectra, 
and the central sharp feature, at 6315 cm$^{-1}$, is at precisely their sum frequency.  
These resonant features in the nonlinear scattering spectrum therefore correspond to 
second-order intra- and inter-mode interactions of the fields, which are resonantly scattered 
into the microcavity modes from the femtosecond (fs) pulses, as the energy in the cavity decays 
due to out-of-plane (linear) scattering.  The sum-frequency generation (SFG) associated with 
the inter-mode interaction provides a weak probe of the mode occupation(s).  
This signal is only non-zero when two modes are excited in the microcavity, and it
may therefore be applicable in weak quantum measurement schemes~\cite{Resch, Pryde}.  

\begin{figure}[h]
\centering
\includegraphics[width=12cm]{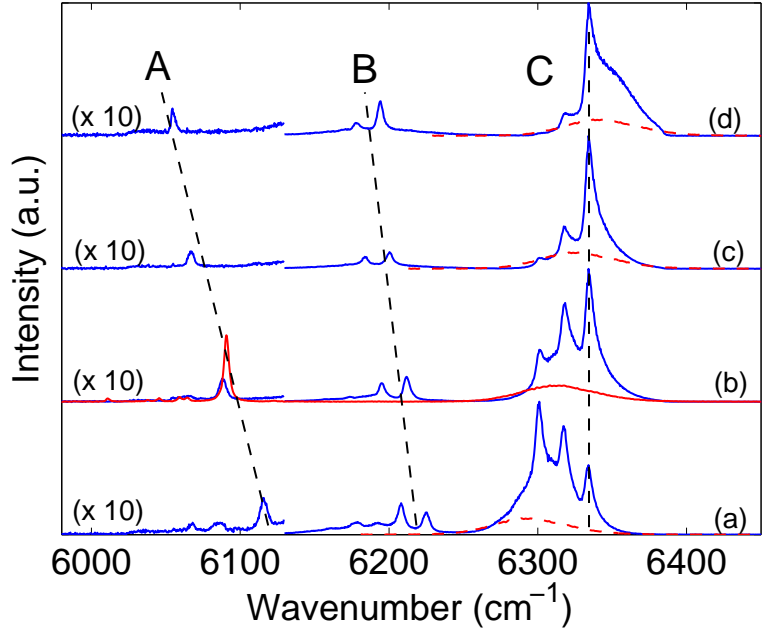}
\caption{\label{fig:idler}Spectra obtained from the interaction of a non-resonant,
narrow linewidth idler pulse; a broad resonant laser pulse; and a two-mode PPC microcavity. 
The blue curves (a)-(d) show the second-order response, plotted at half the energy, as the
OPO is tuned to higher energy.  In each spectrum, the amplitude of the low energy region 
has been multiplied by 10 for clarity.  The solid red curve (b) shows the 
laser spectrum scattered from an untextured part of the InP slab.
The non-resonant second-order laser background is shown 
schematically by the dashed red curves.  The features A, B, and C are discussed in detail in
the text.}
\end{figure}

The electric fields trapped in the microcavity modes can also be used to generate second-order 
radiation at different frequencies via sum-frequency mixing with a separate field incident on 
the structure.  This is illustrated in Figure~\ref{fig:idler}, where a series of four 
second-order spectra from the same microcavity as in Fig.~\ref{fig:2mode} 
are shown when simultaneously excited by short, resonant pulses (as in Fig.~\ref{fig:2mode}), 
and longer, picosecond (ps) pulses tuned far off 
resonance with the microcavity modes.  This two-colour source is readily available from the
 unfiltered ``signal'' beam output of the OPO when it is tuned near the degeneracy point 
(where both signal and idler frequencies are close to half the pump frequency).  
An example of this unfiltered spectrum when the signal is tuned to 6320 cm$^{-1}$ is shown as 
the solid red curve (b) in Figure~\ref{fig:idler}. The short OPO signal pulses are accompanied by 
relatively long (a few ps) OPO idler pulses at a centre frequency roughly equal to the difference 
between the pump and signal frequencies.  The centre frequency of these ps pulses 
converges with the signal beam frequency at half the pump frequency as the OPO is tuned.

The second-order spectra show three principal groups of features that are marked by lines A, B, 
and C to guide the eye. Feature A and the broad background in group C (the fit of which is 
plotted separately as a dashed red line), are the second-harmonics of the ps and 
fs features, respectively, in the excitation spectra.  These features, which shift
at the same rate as the corresponding features in the excitation spectra, are due
to non-resonant second-order scattering, as observed in Fig.~\ref{fig:highQ}. 

The three sharp features in group 
C that do not shift, and the two sharp features in group B that shift at half the rate 
of the excitation beam(s), are specific to the microcavity modes.  They all reflect 
second-order processes that involve the fields ``trapped'' in at least one of the modes, 
as they ring down.  The three (fixed) peaks in group C correspond to the mode SHG and 
SFG features, as in Figure~\ref{fig:2mode}.  The features in group B are then easily
understood to result from the second-order radiation of the two field distributions 
trapped in the microcavity modes respectively interacting with the ps pulses that are 
constantly irradiating the cavity during the ring-down.  To understand the difference
in the shift rate between features A and B, consider a ps pulse at $\omega_A$ interacting
with a microcavity mode at $\omega_C$.  When the ps pulse is tuned from $\omega_A$
to $\omega_A - \Delta\omega$, the second-order feature A shifts from 2$\omega_A$ to
$2(\omega_A-\Delta\omega)$, which is a shift of $-2\Delta\omega$, whereas feature B shifts
from $\omega_A+\omega_C$ to $\omega_A - \Delta\omega +\omega_C$, a shift of just
$-\Delta\omega$.  The processes illustrated here demonstrate that the fields stored in
microcavity modes can be used in conjunction with tunable sources to produce 
second-order radiation in a spectral window of choice.  

Experimentally, the polarization of the non-resonant second-order radiation associated
with the laser background in spectra such as shown in Fig.~\ref{fig:highQ}
is virtually unchanged from spectra obtained from an untextured ``bulk'' part of the InP
slab.  This suggests that the bulk InP second-order tensor is sufficient to describe the 
nonlinear polarization generated in the material by the OPO pulses.  However,
the relationship between the {\em electromagnetic
field polarizations} of the excitation source and the second-order fields radiated from
the resonant modes is
quite different, and more complex, in this microcavity geometry than in more
familiar bulk or uniform waveguide structures.  The second-order polarizations associated 
with the resonant modes have complex multipole-like distributions, and scattering from 
the surrounding lattice of air-holes plays an important role in determining the properties
of the far-field radiation that is generated.  An analysis of the polarization properties 
of the second-order radiation will be discussed in a forthcoming publication.

\section{Summary}

In summary, we have demonstrated second-order intra- and inter-mode nonlinear mixing of 
3D-localized modes in a planar photonic crystal defect microcavity.  When the 
microcavity supports a single mode, the second-order spectrum mimics the linear 
spectrum, showing both non-resonant and resonant features.  When the microcavity 
supports two modes, an additional feature is revealed in the second-order spectrum, due 
to the nonlinear mixing of both resonant modes.  The energy stored in the microcavity 
modes can be used in conjuction with a separate, nonresonant beam, to generate tunable 
sum frequency radiation over a broad range of frequencies.

\section*{Acknowledgements}

The authors  wish to  acknowledge the financial support of the Natural Sciences and
Engineering Research Council of Canada, the Canadian Institute for Advanced Research,
the Canadian Foundation for Innovation, the Canadian Institute for Photonic Innovations, 
and the technical assistance of Lumerical Solutions~Inc.

\end{document}